\documentclass{article}

\PassOptionsToPackage{numbers, compress}{natbib}
\usepackage[final]{neurips_2023}

\usepackage[utf8]{inputenc} % allow utf-8 input
\usepackage[T1]{fontenc}    % use 8-bit T1 fonts
\usepackage{hyperref}       % hyperlinks
\usepackage{url}            % simple URL typesetting
\usepackage{booktabs}       % professional-quality tables
\usepackage{amsfonts}       % blackboard math symbols
\usepackage{nicefrac}       % compact symbols for 1/2, etc.
\usepackage{microtype}      % microtypography
\usepackage{xcolor}         % colors
\usepackage{graphicx}
\usepackage{algorithm}
\usepackage{algorithmic}
\usepackage{multirow}
\usepackage{xspace}

\newcommand{\name}{{\textsc{SimKGCL}}\xspace}
\newcommand{\eat}[1]{}

\title{On the Sweet Spot of Contrastive Views for Knowledge-enhanced Recommendation}

\begin{document}

\maketitle

\begin{abstract}
  %In recommender systems, knowledge graph (KG) can provide additional knowledge that is lacking in the original user-item interaction graph (IG). Recently, graph neural networks (GNNs) based model has gradually become the theme of knowledge-enhanced recommendation. However, most of existing methods mainly consider exploiting the connectivity of entities in KG, while neglecting the critical effect of multi-hop neighborhood information in IG. Moreover, sparse supervised signal and insufficient knowledge incorporation have limited the further performance improvement. To address above issues, we propose a novel knowledge-enhanced model architecture, which explores the high-order connections in both KG and IG, and incorporates KG embeddings into the message aggregation process of IG to make full use of valuable knowledge information. Furthermore, a cross-views constrastive learning framework is proposed to encode user/item representations shared by both IG and KG views. Specifically, we maximize the mutual information between the final predictive representations and knowledge-aware representations, which ensures that the incorporated KG information is useful to the recommendation task. Extensive experimental results on three real-world datasets demonstrate the effectiveness and efficiency of our method, compared to state-of-the-art methods.

  In recommender systems, knowledge graph (KG) can offer critical information that is lacking in the original user-item interaction graph (IG). Recent process has explored this direction and shows that contrastive learning is a promising way to integrate both.
  However, we observe that existing KG-enhanced recommenders struggle in balancing between the two contrastive views of IG and KG, making them sometimes even less effective than simply applying contrastive learning on IG without using KG.
  In this paper, we propose a new contrastive learning framework for KG-enhanced recommendation. Specifically, to make full use of the knowledge, we construct two separate contrastive views for KG and IG, and maximize their mutual information; to ease the contrastive learning on the two views, we further fuse KG information into IG in a one-direction manner.
  Extensive experimental results on three real-world datasets demonstrate the effectiveness and efficiency of our method, compared to the state-of-the-art. Our code is available through the anonymous link: \url{https://figshare.com/articles/conference\_contribution/SimKGCL/22783382}.

  % Our code is released at https://github.com/libuyan-nuaa/simkgcl.
%\yy{I feel that the sayings from Knowledge-Adaptive Contrastive Learning for Recommendation are quit a fit for your work; we may also think it this way: that is, KG contains many noisy information which could hurt the recommendation performance if it is not clearly used;}
\end{abstract}

%%
%% This is file `001intor.tex',  the introduction part of the paper
\section{Introduction}
The success of recommender systems has made them popular in many real-world applications for predicting personalized preferences~\cite{ricci2011introduction}. Traditional recommendation methods that solely rely on historical user-item interactions (e.g., collaborative filtering based ones~\cite{he2017neural2,rendle2012bpr,wang2019neural,lin2022improving}) severely suffer from data sparsity and cold-start problems, %as they often treat each interaction as an independent instance without considering the relationships between them. 
%To overcome this limitation, 
and thus knowledge graph (KG) has been leveraged to encode additional item semantic relevance~\cite{cao2019unifying,wang2018ripplenet,tai2020mvin,wang2019kgat,wang2019knowledge1,wang2019knowledge,wang2021learning}.

\begin{figure}[t]
  \centering
  \includegraphics[width=\linewidth]{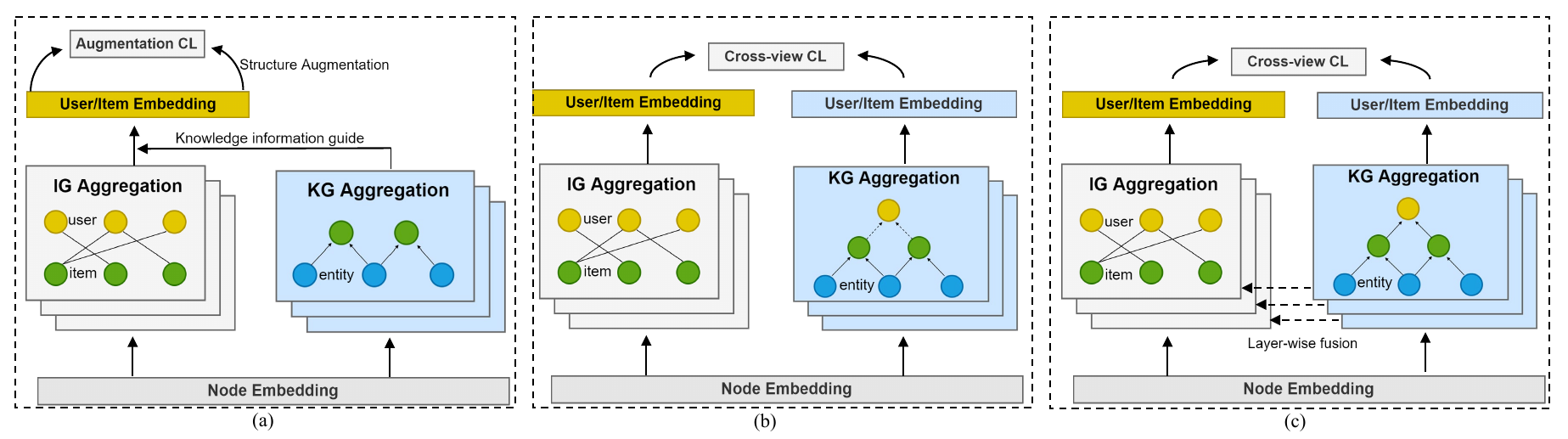}
  \caption{Conceptual comparison between existing KCL methods (a and b) and ours (c).}
  \label{fig: model_comparision}
  % \Description{Figure 1: fig1.}
\end{figure}

Among existing KG-enhanced methods, encoding KG with graph neural networks (GNNs) and further connecting it to the user-item interaction graph (IG) with contrastive learning (CL) has been the recent trend and shown prominent performance (see the related work section for more details).
However, we observe that existing recommenders that are built on knowledge graph and contrastive learning still have room for improvement. 
For example, they are even less effective compared to the methods solely applying CL on IG without using KG information.
% For example, they are sometimes even less effective than simply applying contrastive learning on IG without using KG information. 
We name the existing work as {\em knowledge-enhanced contrastive learning (KCL)} methods, and summarize them in Figure~\ref{fig: model_comparision}. 

%We revisit the recently proposed knowledge-enhanced contrastive learning (KCL) methods and find them fall short in fully incorporating knowledge information and constructing appropriate CL views. 
Specifically, existing KCL methods can be divided into two categories.
Figure~\ref{fig: model_comparision}(a) shows the first category where CL is conducted based on IG's data augmentation and KG is used to guide the augmentation. For example, KGCL~\cite{yang2022knowledge} encodes item embeddings with KG to serve as the initialization in IG, and uses knowledge to guide the structure augmentation of IG for constructing CL views. %Though knowledge information is used, Its manner of generating CL views with single IG is too ``easy'' to learn valuable information shared by both KG and IG. 
However, according to the ``InfoMin principle''~\cite{tian2020makes}, i.e., good CL views are those that share the minimal information necessary for the downstream, it is desired to generate CL views from both KG and IG to obtain shared representations of users and items. We name the issue of Figure~\ref{fig: model_comparision}(a) as {\em excessive overlap}. 

Figure~\ref{fig: model_comparision}(b) shows the second category, where existing KCL methods (e.g.,~\cite{zou2022multi,wang2023knowledge,pan2021collaborative}) perform cross-view CL between KG and IG.
%, and integrate (e.g., sum or concatenation) both embeddings for recommendation. 
However, they separately encode KG and IG, leading to insufficient communication between the two views. As indicated by the ``modality gap'' phenomenon~\cite{liang2022mind}, embeddings from IG and KG are potentially located in two completely separate regions of the embedding space, causing significant burdens for CL to determine the shared representations.
%which indicates that it may be too ``hard'' to directly align KG and IG embeddings. 
We name this issue as {\em insufficient overlap}.

To address the issues of existing work, we propose a simple but effective KCL framework, named \name. As shown in Figure~\ref{fig: model_comparision}(c), we build CL views from both IG and KG to mitigate the excessive overlap issue; we further fuse the KG embeddings into the encoding process of IG in a one-direction, layer-wise manner, to mitigate the insufficient overlap issue. %n, to make full use of knowledge information and avoid the gap between IG and KG views of CL task.
Extensive experiments on three real-world datasets demonstrate the effectiveness and efficiency of our proposed method. 
Specifically, we compare \name with 16 existing baselines, including all the four existing KCL methods we know, and the results show that \name achieves 6.5\% - 13.2\% relative improvements compared the best competitor among them. 
Further, compared with the state-of-the-art KCL method KGCL~\cite{yang2022knowledge}, \name is much more efficient and requires only $1/30$ training time on largest MIND dataset. 

\section{Related Work}
%In this section, we briefly review the related work, including Knowledge-enhanced recommendation systems and constrastive learning in recommendation.
{\bf Contrastive Learning for Recommendation}.
 Inspired by the progress of self-supervised learning in NLP~\cite{fu2021lrc,devlin2018bert} and CV~\cite{he2020momentum,chen2020simple} domains, several recent studies~\cite{wu2021self,lin2022improving,yu2022xsimgcl,yu2022graph,yang2022knowledge,zou2022multi} have attempted to introduce contrastive learning into recommender systems. %Sepcifically, some methods consider contrastive learning as auiliary task and have achieved remarkable results. 
Among them, SGL~\cite{wu2021self} generates different contrastive views through random structure augmentation (e.g., edge drop); NCL~\cite{lin2022improving} employs a prototypical contrastive objective to capture the correlations between a user/item and its prototype; SimGCL~\cite{yu2022graph} directly adds random uniform noise to the representation for data augmentation.

{\bf Knowledge-enhanced Recommendation}.
Depending on how they encode the KG, knowledge-enhanced recommender systems can be roughly divided into three categories: embedding-based methods~\cite{cao2019unifying,huang2018improving,wang2018shine,wang2018dkn,wang2019multi} which utilize traditional KG embedding techniques (e.g., TransE~\cite{bordes2013translating} and TransH~\cite{wang2014knowledge}), path-based methods~\cite{hu2018leveraging,shi2018heterogeneous,wang2019explainable,yu2013collaborative,yu2014personalized,zhao2017meta} which explore long-range connectivity among items via pre-defined paths in the KG, and GNN-based methods~\cite{wang2019kgat,wang2021learning,wang2019knowledge,wang2019knowledge1,jin2020efficient,wang2020ckan} which leverage the message propagation mechanism of graph neural networks~\cite{wu2020comprehensive}. Among them, embedding-based methods struggle to capture high-order dependencies, path-based methods heavily depend on the quality of paths, and GNN-based methods have shown promising results by addressing the limitations of the former two. However, all these methods follow the supervised learning paradigm, relying on the observed user-item interactions which are usually extremely sparse in real-world scenarios.

{\bf Contrastive Learning for Knowledge-enhanced Recommendation}.
%\yy{let's organize existing work into three classes: KG-based recommendation, CL-based recommendation, and KGCL-based recommendation.}
More recently, researchers have attempted to harness contrastive learning for improving the performance of knowledge-enhanced recommendation~\cite{yang2022knowledge,zou2022multi,wang2023knowledge,zou2022improving,pan2021collaborative}. %self-supervised learning into knowledge-aware recommender systems
 For example, KGCL~\cite{yang2022knowledge} leverages knowledge graph semantics to guide the structure augmentation of CL views. MCCLK~\cite{zou2022multi} and CKER~\cite{pan2021collaborative} introduce a CL module to learn user/item representations shared by both IG and KG. KACL~\cite{wang2023knowledge} further adaptively drops irrelevant edges in KG and IG to construct more reliable CL views. Nevertheless, they still suffer from either the excessive overlap or insufficient overlap issues as discussed in introduction.  %these methods need further improvement(in our experiments, even less effective than CL work without using KG) due to insufficient incorporation of knowledge information. Motivated by existing efforts, in this paper, we propose a simpler and more efficient contrastive learning paradigm for knowledge-enhanced recommendation. 

\section{Problem Statement}
In this section, we state the problem definition. %We first introduce two types of graphs (i.e., interaction graph and knowledge graph) for recommendation, followed by the problem statement of the knowledge-aware recommendation problem.

\textbf{Interaction Graph.} In a recommender system, we have user set $\mathcal{U} = \{u_1,u_2,...,u_M\}$ and item set $\mathcal{I} = \{i_1,i_2,...,i_N\}$, where $M$ and $N$ represent their sizes, respectively. The interaction graph is denoted as $\mathcal{G}_R = \{(u,i,y_{u,i})\}$, where entry $y_{u,i} = 1$ if there exists an observed interaction between user $u$ and item $i$, and $y_{u,i} = 0$ otherwise. %The user-item interactions are denoted as ${\bf Y} \in \{0,1\}^{M \times N}$, 

%In addition to historical interactions, w
\textbf{Knowledge Graph.} We denote the knowledge graph as $\mathcal{G}_K = \{(h,r,t)\ | h,t \in \mathcal{E}, r \in \mathcal{R} \}$, where  $\mathcal{E}$ and $\mathcal{R}$ represent the entity set and relation set, and each triple $(h,r,t)$ describes a relationship between entities. 
%knowledge graph which includes external item knowledge with different types of entities and relations. Specifically, $h,r,t$ denote the head entity, relation and tail entity of a knowledge triplet, respectively; $\mathcal{E}$ and $\mathcal{R}$ represent the entities set and relation set of $\mathcal{G}$.
For example, the triplet ({\em Titanic}, {\em Directed by}, {\em James Cameron}) means that the movie Titanic is directed by James Cameron. 
Note that $\mathcal{E}$ may contain both items and non-items (e.g., {\em Titanic} and {\em James Cameron}, respectively, in a movie recommendation scenario). To connect the two graphs, we establish a set of item-entity alignments $\mathcal{A} = \{ (i, e) | i \in \mathcal{I}, e \in \mathcal{E}\}$, where $(i,e)$ represents that item $i$ can be aligned with entity $e$.

\textbf{Task Formulation.} Given the interaction graph $\mathcal{G}_R$ and knowledge graph $\mathcal{G}_K$ as input, our goal is to learn a score function to forecast the items that user $u$ would like to interact with. %(u \in \mathcal{U})$

\section{The Proposed Approach}
\begin{figure}[t]
  \centering
  \includegraphics[width=\linewidth]{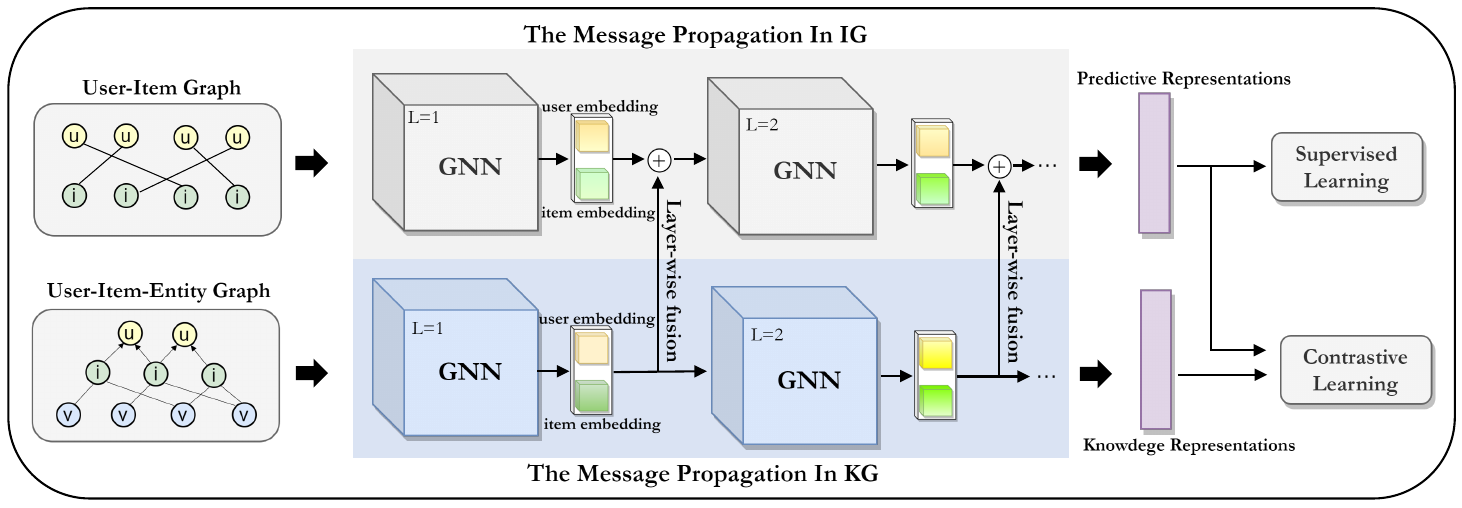}
  \caption{The overview of our \name.}%\yy{also needs revision, you may follow the sigir papers to draw a similar one.} \lxj{to be checked.}
  \label{fig:overview}
  % \Description{Figure 1: fig1.}
\end{figure}

In this section, we present the proposed method \name. 
\eat{In this section, we introduce the overall framework of the model, mainly about how to combine the embeddings of KG part and IG part to obtain the final users/items representation for prediction. As shown in Figure~\ref{fig: model_comparision}, we compare the architecture difference between existing Knowledge graph recommendation(KGR) methods and ours. Specifically, as shown in Figure~\ref{fig: model_comparision}a), most of the previous KGR methods~\cite{wang2018ripplenet,wang2019knowledge,wang2019kgat,wang2021learning} only use knowledge graph to learn better embeddings of the items, and then obtain the embeddings of the users by considering the first-order neighbors(connection of $user \leftarrow item$) in the interaction graph. The whole process does not consider the connections from users to items(i.e., $item \leftarrow user$), and does not utilize the high-order neighbor information in the interaction graph. 
As shown in Figure~\ref{fig: model_comparision}b), some methods like~\cite{yang2022knowledge} learn better items representation through KG encoder as the initial embeddings of information aggregation in IG. We think that the information of KG and IG aggregator should be fully integrated to learn more accurate users/items representation, such that a layer-wise fusion operation is adopted in our model. As shown in Figure~\ref{fig: model_comparision}c), we perform message propagation of KG and IG at the same time, and then integrate KG embedddings into the embedding aggregator of IG in each layer.}
Figure~\ref{fig:overview} shows the overview of \name, which consists of the following four key components: 1) message propagation in KG, 2) message propagation in IG, 3) one-direction layer-wise fusion, and 4) contrastive learning.% \yy{this part needs to be revised}

%\subsection{Separate Message Propagation}
%In this section, we present the overall model design, including the message propagation mechanisms in the knowledge graph(KG) and interaction graph(IG).

\subsection{Message Propagation in KG}
Following the recent trend, we adopt GNNs as the backbone to encode both KG and IG.
For KG, we first define the message propagation for entities and items (i.e., $entity \leftarrow entity$ and $item \leftarrow entity$).
Inspired by~\cite{wang2021learning}, we model the relation-aware message propagation as follows:
\begin{equation}\label{eq:kg_propagate}
%\begin{array}{l}
\mathbf{e}_{i}^{(l)}=\frac{1}{\left|\mathcal{N}_{i}^{K}\right|} \sum\limits_{(r, v) \in \mathcal{N}_{i}^{K}} \mathbf{e}_{r} \odot \mathbf{e}_{v}^{(l-1)}, \quad
\mathbf{e}_{v}^{(l)}=\frac{1}{\left|\mathcal{N}_{v}^{K}\right|} \sum\limits_{(r, v^{'}) \in \mathcal{N}_{v}^{K}} \mathbf{e}_{r} \odot \mathbf{e}_{v^{'}}^{(l-1)},
%\end{array}
\end{equation}	
where $\mathcal{N}_{i}^{K}/$$\mathcal{N}_{v}^{K}$ represents the one-hop entity neighbors of item $i/$entity $v$ in the KG, $l \in \{1, 2, ..., L\}$ is the propagation depth, and $\mathbf{e}_{i}^{(l)}/$$\mathbf{e}_{v}^{(l)}$ denotes the embedding of item $i/$entity $v$ in layer $l$. 
For each triplet $(i,r,v)$, we obtain a relational message $\mathbf{e}_{r} \odot \mathbf{e}_{v}^{(l)}$ by modeling the relation $r$ as a rotation  operator~\cite{sun2019rotate} using element-wise product. Note that the element-wise product may result in significantly large numbers, and thus we conduct a normalization step in each propagation layer.

%\yy{why users in KG? is it because the new KG is actually constructed to combine both the original KG and IG?}
Then, to make the learned embeddings compatible with the IG, we also borrow the user-item interactions to obtain the message propagation for users (i.e., $user \leftarrow item \leftarrow entity$):
%by considering one-hop neighbors in the interaction graph:
\begin{equation} \label{eq:kg_item2user}
\mathbf{e}_{u}^{(l)}=\sum_{i \in \mathcal{N}_{u}^{I}} \frac{1}{\sqrt{\left|\mathcal{N}_{u}^{I}\right|\left|\mathcal{N}_{i}^{I}\right|}} \mathbf{e}_{i}^{(l)},
\end{equation}
where $\mathcal{N}_{u}^{I}/$ $\mathcal{N}_{i}^{I}$ represents the one-hop item/user neighbors for user $u$/item $i$ in the IG. Note that we perform message propagation $user \leftarrow item$ after $item \leftarrow enity$ such that the information of $\mathbf{e}_{u}^{l}$ essentially comes from entities in the KG.

%The superscripts of user embedding and item embedding are both $l$, indicating that we perform the information aggregation $user \leftarrow item$ after $item \leftarrow enity$ such that the information of $\mathbf{e}_{u}^{l}$ essentially comes from entities in knowledge graph $\mathcal{G}$.

\eat{
It is noting that when the embedding elements are greater than 1, the element-wise product operation in Eq.~\ref{eq:kg_propagate} will make the elements in embedding increase continuously. In order to avoid the influence of excessive element values in embedding on model convergence, we conduct a normalize operation in each aggregation layer:
\begin{equation} \label{eq: normalize}
\mathbf{e}_{i}^{(l)} = \frac{\mathbf{e}_{i}^{(l)}} {\left\|\mathbf{e}_{i}^{(l)}\right\|_{2}},
\mathbf{e}_{u}^{(l)} = \frac{\mathbf{e}_{u}^{(l)}} {\left\|\mathbf{e}_{u}^{(l)}\right\|_{2}},
\end{equation}
}

\subsection{Message Propagation in IG}
%As a fundamental component in recommender systems, collaborative filtering (CF) aims to model user preferences based on observed feedback. 
For IG, we encode user-item interactions by utilizing neural graph collaborative filtering methods. Inspired by the effective and lightweight architecture of LightGCN~\cite{he2020lightgcn}, we discard the feature transformation and the non-linear activation layer in the aggregation function, and conduct message propagation as:
\begin{equation}\label{eq:ig_propagate}
%\begin{array}{l}
\mathbf{x}_{u}^{(l+1)}=\sum_{i \in \mathcal{N}_{u}^{I}} \frac{1}{\sqrt{\left|\mathcal{N}_{u}^{I}\right|\left|\mathcal{N}_{i}^{I}\right|}} \mathbf{x}_{i}^{(l)}, \quad
\mathbf{x}_{i}^{(l+1)}=\sum_{u \in \mathcal{N}_{i}^{I}} \frac{1}{\sqrt{\left|\mathcal{N}_{i}^{I}\right|\left|\mathcal{N}_{u}^{I}\right|}} \mathbf{x}_{u}^{(l)},
%\end{array}
\end{equation}	
where $\mathbf{x}_{u}^{(l)}/$$\mathbf{x}_{i}^{(l)}$ denotes the aggregated embedding of user $u/$item $i$ in layer $l$. %\yy{we have two $N_i$'s. can we somehow distinguish them?} \lxj{we use $\mathcal{N}_{u}^{I}/$ $\mathcal{N}_{i}^{I}$ denotes neighbors in IG, and use $\mathcal{N}_{i}^{K}$/$\mathcal{N}_{v}^{K}$ denotes neighbors in KG.}

\subsection{One-direction Layer-wise Fusion}
% \yy{In this section, we introduce the overall framework of the model, mainly about how to combine the embeddings of KG part and IG part to obtain the final users/items representation for prediction. As shown in Figure~\ref{fig: model_comparision}, we compare the architecture difference between existing Knowledge graph recommendation(KGR) methods and ours. Specifically, as shown in Figure~\ref{fig: model_comparision}a), most of the previous KGR methods~\cite{wang2018ripplenet,wang2019knowledge,wang2019kgat,wang2021learning} only use knowledge graph to learn better embeddings of the items, and then obtain the embeddings of the users by considering the first-order neighbors(connection of $user \leftarrow item$) in the interaction graph. The whole process does not consider the connections from users to items(i.e., $item \leftarrow user$), and does not utilize the high-order neighbor information in the interaction graph. 
% As shown in Figure~\ref{fig: model_comparision}b), some methods like~\cite{yang2022knowledge} learn better items representation through KG encoder as the initial embeddings of information aggregation in IG. We think that the information of KG and IG aggregator should be fully integrated to learn more accurate users/items representation, such that a layer-wise fusion operation is adopted in our model. }

%\yy{the whole discussion above should be placed earlier, either/both in intro and the begnining of section 4. we may simply mention a few words as I listed below.} \lxj{I have stated it in intro.}

As shown in Figure~\ref{fig: model_comparision}, one of the key designs of our method is to perform one-direction layer-wise fusion from KG to IG.
%Existing KCL studies~\cite{yang2022knowledge,zou2022multi,wang2023knowledge,pan2021collaborative} do not incorporate knowledge into the message propagation process of IG, and directly contrast KG and IG embddings may fall short in modality gap problem.  
%To overcome these issues, we introduce an one-direction layer-wise fusion operation to fully integrate both, and avoid the gap of CL views. 
One may consider the two-way message propagation between KG and IG; however, this would collapse the two views into one, violating the ``InfoMin principle''.\footnote{This is also confirmed by our experiments.}
%\yy{fill in the above two sentences.} \lxj{to be checked.}
Specifically, since we use GNNs as the backbone, we incorporate the learned KG embeddings into IG in each propagation layer:
\begin{equation}\label{eq:fusion}
%\begin{array}{l}
\mathbf{x}_{u}^{(l)}=\mathbf{x}_{u}^{(l)} + \mathbf{e}_{u}^{(l)}, \quad
\mathbf{x}_{i}^{(l)}=\mathbf{x}_{i}^{(l)} + \mathbf{e}_{i}^{(l)}.
%\end{array}
\end{equation}
%where $\mathbf{x}_{u}^{(l)}, \mathbf{x}_{i}^{(l)}$ represent the aggregation embedding in IG of user $u$ and item $i$, respetively; $\mathbf{x}_{u}^{(l)}, \mathbf{x}_{i}^{(l)}$ stand for the aggregation embedding in KG of user $u$ and item $i$, respectively. 
After propagating $L$ layers, we utilize the weighted sum function as the readout function to obtain final predictive representations $\mathbf{x}_u$ and $\mathbf{x}_i$ of users and items, respectively, %which combines the fused embeddings of all layers as follows:
\begin{equation} \label{eq: final_representation}
\mathbf{x}_{u}=\frac{1}{L} \sum_{l=1}^{L} \mathbf{x}_{u}^{(l)}, \quad \mathbf{x}_{i}=\frac{1}{L} \sum_{l=1}^{L} \mathbf{x}_{i}^{(l)}.
\end{equation}
%where $\mathbf{x}_u$ and $\mathbf{x}_i$ denote the predictive representations of user $u$ and item $i$, respectively. 

\eat{Then, the inner product between them is calculated to predict the preference score of user $u$ towards item $i$:

\begin{equation}\label{eq:prediction_score}
\hat{y}_{u, i}=\mathbf{x}_{u}^{\top} \mathbf{x}_{i}.
\end{equation}
}

We then adopt the Bayesian Personalization Ranking (BPR) loss~\cite{rendle2012bpr} to serve the supervised learning task. Specifically, we sample one negative sample from unobserved user-item pairs for each observed user-item interaction to construct the pairwise training data, and obtain the following objective function:
\begin{equation}\label{eq:bpr}
\mathcal{L}_{BPR}=\sum_{(u, i, j) \in \mathcal{O}}-\log \sigma\left(\hat{y}_{u i}-\hat{y}_{u j}\right),
\end{equation}
where $\mathcal{O}=\left\{(u, i, j) \mid u \in \mathcal{U}, i \in \mathcal{I}, j \in \mathcal{I},{y}_{u,i} \neq 0, {y}_{u,j}=0\right\}$,  the prediction score $\hat{y}_{u, i}$ is computed as the inner-product between $\mathbf{x}_u$ and $\mathbf{x}_i$, and $\sigma$ denotes the sigmoid function. %BPR loss promotes the observed user-item interactions to obtain higher prediction scores than the unobserved ones.

\subsection{Cross-view Contrastive Learning}
%\lxj{I have revised this subsection, to be checked.}
%Inspired by the recent contrastive learning studies~\cite{wu2021self,yu2022graph,yang2022knowledge, zou2022multi,wang2023knowledge,pan2021collaborative}, we expect to learn better user/item representations based on Mutual Information Maximization (MIM)~\cite{peng2020graph, velickovic2019deep} among different views. Unlike previous methods that construct different CL views by conducting structure augmentation~\cite{wu2021self, yang2022knowledge} on single IG, we apply cross view CL to learn representations shared by both KG and IG, and avoid embedding gap with help of layer-wise fusion.
%we propose to use both the fused IG (i.e., Eq.~\ref{eq: final_representation}) and the KG as the two contrastive views.
%we design a simple and effective comparative learning framework by directly generating different views with fused IG embeddings and single KG embeddings.
\eat{
On the one hand, we regard the final representation in Eq~\ref{eq: final_representation} as the first perspective of contrastive learning, which has considered the aggregation embeddings in both IG and KG. On the other hand, we generating KG representations as another view by only utilizing the embedding in KG:}
%Specifically, w
Our contrastive learning maximizes the mutual information between predictive representations from IG (i.e., Eq.~\ref{eq: final_representation}) and the knowledge-aware representations from KG. 
%both of them derive from the fused IG embeddings and KG embeddings, respectively. 
The knowledge-aware representations of user $u$ and item $i$ are calculated as follows, 
\begin{equation} \label{eq: kg_presentation}
\mathbf{e}_{u}=\frac{1}{L+1} \sum_{l=0}^{L} \mathbf{e}_{u}^{(l)}, \quad 
\mathbf{e}_{i}=\frac{1}{L+1} \sum_{l=0}^{L} \mathbf{e}_{i}^{(l)}.
\end{equation}
where we also use a weighted-sum readout function on the KG embeddings.
%final predictive representations

% \yy{put it somewhere else: The most similar to our method is MCCLK~\cite{zou2022multi}, the main difference is that we adopt layer-wise fusion, and our design is simpler and more efficient.}
Then, we adopt InfoNCE loss~\cite{van2018representation} to optimize a lower bound of mutual information, which can be written as follows:
\begin{equation}\label{eq:cl_user}
\mathcal{L}_{cl}^{user}=\sum_{u \in \mathcal{U}}-\log \frac{\exp \left(\cos\left(\mathbf{x}_{u}, \mathbf{e}_{u}\right) / \tau\right)}{\sum_{j \in \mathcal{U}} \exp \left(\cos\left(\mathbf{x}_{u}, \mathbf{e}_{j}\right) / \tau\right)},
\end{equation} 
where $\tau$ is the hyper-parameter, also known as the \textit{temperature} in softmax, and $\cos(\cdot{})$ denotes the cosine similarity function. Analogously, we can obtain the InfoNCE loss for the item side as $\mathcal{L}_{cl}^{item}$. Hence, we have the final contrastive learning loss function
%combine the above two loss functions to form the objective function of the contrastive learning task as: 
%\begin{equation}\label{eq:cl_loss}
$\mathcal{L}_{cl} = \mathcal{L}_{cl}^{user} + \mathcal{L}_{cl}^{item}$.
%\end{equation}

%\subsection{Model Training and Analysis}
%\subsubsection{Multi-task Training}
%In order to learn the recommendation task and the self-supervised task at the same time, we jointly optimize the whole model with a multi-task training strategy. 
Overall, the training loss of our method is
\begin{equation}\label{eq:all_loss}
\mathcal{L}=\mathcal{L}_{\mathrm{BPR}}+\lambda_{1} \mathcal{L}_{\mathrm{cl}}+\lambda_{2}\|\Theta\|_{F}^{2},
\end{equation}
where $\lambda_{1}$ and $\lambda_{2}$ are hyper-parameters to control the strengths of the contrastive loss and regularization term, respectively, and $\Theta$ denotes the learnable model parameters. 

\subsection{Complexity Analysis}
\begin{table}\centering
	\caption{The comparison of time complexity.}
	\label{tab:time}
	\begin{tabular}{cccc}
		\toprule
		Component & LightGCN & KGCL & \name\\
		\midrule
%		Adjacency Matrix & $\mathcal{O}(2|\mathcal{G}_R|)$ & $\mathcal{O}(2|\mathcal{G}_R|+4 \rho|\mathcal{G}_R|)$ &$\mathcal{O}(2|\mathcal{G}_R|)$\\ 
		IG encoding & $\mathcal{O}(2|\mathcal{G}_R| Ld)$ & $\mathcal{O}((2+4 \rho)|\mathcal{G}_R| Ld)$ & $\mathcal{O}(2|\mathcal{G}_R| Ld)$\\
		KG encoding & - & $\mathcal{O}((2+4 \rho')|\mathcal{G}_K| Ld)$ & $\mathcal{O}(2|\mathcal{G}_K| Ld)$\\
            BPR loss computing & $\mathcal{O}(2Bd)$ & $\mathcal{O}(2Bd)$ & $\mathcal{O}(2Bd)$\\
		CL loss computing & - & $\mathcal{O}(BMd)$ & $\mathcal{O}(BMd)$\\
		\bottomrule
	\end{tabular}
\end{table}

%\textbf{Model Size}. 
Since we introduce no additional trainable parameters, our space complexity is the same with baselines~\cite{he2020lightgcn,wang2019kgat}. 
%\textbf{Time Complexity}. 
The time cost of \name mainly comes from the message propagation in IG and KG, BPR loss computing, and contrastive learning loss computing, which keeps the same order of magnitude with the lightweight baseline LightGCN~\cite{he2020lightgcn}. Table~\ref{tab:time} presents the time complexity results of LightGCN~\cite{he2020lightgcn}, KGCL~\cite{yang2022knowledge}, and our \name. In the table, we let $|\mathcal{G}_R|$ be the edge number in IG, $|\mathcal{G}_K|$ be triplet number in KG, $L$ denote the aggregation layers, $d$ denote the embedding size, $B$ denote the batch size, $M$ represent the node number in a batch, and $\rho$ and $\rho'$ denote the edge keep rate and triplet keep rate in KGCL, respectively.
\section{Experiments}
\eat{In this section, we present the experimental results. The experiments are mainly designed to answer the following questions:
\begin{itemize}
    \item {\bf RQ1.} How does the proposed approach perform compared with existing methods?
    \item {\bf RQ2.} How does each component of \name contribute to the overall performance?
    \item {\bf RQ3.} How stable is the proposed method w.r.t. the hyper-parameters?
    \item {\bf RQ4.} How efficient is the proposed method in the training stage?
\end{itemize}
}

\subsection{Experimental Setup}
{\bf Datasets}.
To evaluate the effectiveness of the proposed \name, we conduct extensive experiments on three public datasets gathered from various real-life platforms: Yelp2018, Amazon-book, and MIND. Table~\ref{tab:datasets} presents the datasets' statistics. Fowllowing~\cite{wang2019kgat, yang2022knowledge}, we construct knowledge graphs for Yelp2018 and Amazon-Book datasets by mapping items into Freebase entities~\cite{zhao2019kb4rec}. In our experiments, we only collect entities within two hops since few of methods consider modeling multi-hop relations in KG. For the MIND dataset, we follow the strategy in ~\cite{tian2021joint} to construct the knowledge graph based on spacy-entity-linker tool$\footnote{https://github.com/egerber/spaCy-entity-linker}$ and Wikidata$\footnote{https://query.wikidata.org/}$. For the pre-processing of all datasets, we are consistent with KGCL~\cite{yang2022knowledge}. %\yy{now that KGCL is not the only competitor, is this setup also consistent with others?} \lxj{our all datasets settings is the same with KGCL, not for others.}

\begin{table} \centering
  \caption{Statistics of datasets.}
  \label{tab:datasets}
  \resizebox{1\columnwidth}{!}{
  \begin{tabular}{cccccccc}
    \toprule
    Datasets & \#Users & \#Items & \#Interactions & Density & \#Relations & \#Entities & \#Triplets\\
    \midrule
    Yelp2018 & 45,919 & 45,538 & 1,183,610 & $5.7 \times 10^{-4}$ & 42 & 47,472 & 869,603\\ 
    Amazon-book & 70,679 & 24,915 & 846,434 & $4.8 \times 10^{-4}$ & 39 & 29,714 & 686,516\\
    MIND & 300,000 & 48,947 & 2,545,327 & $1.7 \times 10^{-4}$ & 90 & 106,500 & 746,270\\
  \bottomrule
\end{tabular}
}
\end{table}

{\bf Compared Methods}.
We compare the proposed method with the following 16 methods.
\begin{itemize}
\item \textbf{BPR}~\cite{rendle2012bpr} is a representative recommendation approach to rank item candidates with a pairwise ranking loss.
\item \textbf{NCF}~\cite{he2017neural} utilizes the multiple-layer perceptron to endow the collaborative filtering (CF) architecture with the non-linear feature interaction.
\item \textbf{NGCF}~\cite{wang2019neural} is a graph-based CF method that largely follows the standard GCN, which additionally encodes the second-order feature interaction into message propagation.
\item \textbf{LightGCN}~\cite{he2020lightgcn} devises a light-weight graph convolution without feature transformation and non-linear activation, which is more efficient.
\item \textbf{CKE}~\cite{zhang2016collaborative} adopts TransR to encode the items’ semantic information and further incorporates it into the denoising auto-encoders for learning item representations with knowledge base.
\item \textbf{RippleNet}~\cite{wang2018ripplenet} propagates user preference over KG along with the constructed paths rooted at certain users. It is a memory-like neural model to improve user representations.
\item \textbf{KGCN}~\cite{wang2019knowledge} is a GNN-based method, which integrates multi-hop neighborhood information in KG to enrich item embeddings.
\item \textbf{KGAT}~\cite{wang2019kgat} designs an attentive message passing scheme over the collaborative knowledge graph (CKG) for representation learning.
\item \textbf{CKAN}~\cite{wang2020ckan} extends KGCN by considering the multi-hop neighborhood information in IG, which utilizes different message propagation strategies on IG and KG, respectively.
\item \textbf{KGIN}~\cite{wang2021learning} is a recent GNN-based method to identify latent intention of users, and further performs the relational path-aware aggregation for both user-intent-item and KG triplets.
\item \textbf{SGL}~\cite{wu2021self} introduces CL task into graph-based CF to learn self-supervised signals, which performs random structure augmentation on IG to generate different CL views. 
% We implement SGL-ED which is the suggested version.
\item \textbf{SimGCL}~\cite{yu2022graph} is a newly proposed graph-based CF method, which builds CL views by adding random noises in the embedding space and has made a state-of-the-art performance.
\item \textbf{KGCL}~\cite{yang2022knowledge} is a recent KCL method, which attempts to explore KG semantics and conduct a KG-aware structure augmentation to construct CL views.
\item \textbf{MCCLK}~\cite{zou2022multi} is another recent KCL method, which considers learning user/item representations from multiple views and applying cross view CL as an auxiliary task.
\item \textbf{CKER}~\cite{pan2021collaborative} introduces self-supervised learning to maximize the mutual information between the interaction-side and knowledge-side user preferences.
\item \textbf{KACL}~\cite{wang2023knowledge} is a newly proposed KCL method that generates adaptive KG and IG views to perform cross-view CL and alleviates the interaction domination and task-irrelevant noises.
\end{itemize}
Among the above baselines, the former four are basic recommenders (from BPR to LightGCN), the next six are knowledge-enhanced recommenders (from CKE to KGIN), the next two are CL-based recommenders (SGL and SimGCL), and the final four are KCL recommenders (from KGCL to KACL).

{\bf Evaluation Metrics}.
We evaluate the top-N recommendation performance by using two widely used metrics $Recall@N$ and $NDCG@N$, where $N$ is set to 20 for consistency. Following~\cite{he2020lightgcn, wu2021self}, we adopt the full-ranking strategy~\cite{zhao2020revisiting}, which ranks all the candidate items that the user has not interacted with.

{\bf Implementations}.
For all the compared models, we either use the source code provided by their authors (if it exists) or implement them ourselves with RecBole~\cite{zhao2021recbole}, which is a unified open-source framework for developing and replicating recommendation algorithms. 
To ensure fairness, we fix the embedding size and batch size to 64 and 2048, respectively. We optimize all the models by using Adam optimizer. For each compared method, we refer to the best hyper-parameters in their paper and then fine-tune them carefully so as to achieve the best results we can have. 
For the proposed \name, we fix $\lambda_{2}$ to $1e-4$ which is the same with the compared models. We tune the hyper-parameter $\tau$ in $\{0.1,0.2,0.5,0.8,1.0\}$, and $\lambda_{1}$ in $\{0.01,0.05,0.1,0.2,0.5,1.0\}$ on a validation set.

\begin{table}[t]\centering
  \caption{Effectiveness comparison results. Our \name significantly outperforms the existing methods. The best results are in bold and the second-best are underlined. `--' indicates the method cannot return results within 48 hours. The relative improvements compared to the second-best are also shown.}
  \label{tab:performance}
  \begin{tabular}{ccccccc}
  \toprule
    \multirow{2}{*}{Method}
    & \multicolumn{2}{c}{Yelp2018} & \multicolumn{2}{c}{Amazon-Book} & \multicolumn{2}{c}{MIND}\\ 
    & Recall & NDCG & Recall & NDCG & Recall & NDCG \\ 
    \midrule
    BPR & 0.0555 & 0.0375 & 0.1244 & 0.0658 & 0.0938 & 0.0469\\
    NCF & 0.0535 & 0.0346 & 0.1033 & 0.0532 & 0.0893 & 0.0436\\
    NGCF & 0.0587 & 0.0376 & 0.1139 & 0.0574 & 0.0961 & 0.0483\\
    LightGCN & 0.0682 & 0.0443 & 0.1398 & 0.0736 & 0.1033 & 0.0520\\ \midrule
    CKE & 0.0686 & 0.0431 & 0.1375 & 0.0685 & 0.0901 & 0.0382\\
    RippleNet & 0.0422 & 0.0251 & 0.1058 & 0.0549 & 0.0858 & 0.0407\\
    KGCN & 0.0532 & 0.0338 & 0.1111 & 0.0569 & 0.0887 & 0.0431\\
    KGAT & 0.0675 & 0.0432 & 0.1390 & 0.0739 & 0.0907 & 0.0442\\
    CKAN & 0.0689 & 0.0441 & 0.1380 & 0.0726 & 0.0991 & 0.0499\\
    KGIN & 0.0712 & 0.0462 & 0.1436 & 0.0748 & 0.1044 & 0.0527\\ \midrule
    SGL & 0.0719 & 0.0475 & 0.1445 & 0.0766 & 0.1032 & 0.0539\\
    SimGCL & \underline{0.0811}	& \underline{0.0530} & 0.1579 & 0.0848 & \underline{0.1111} & \underline{0.0594} \\ \midrule
    KGCL & 0.0756 & 0.0493 & 0.1496 & 0.0793 & 0.1073 & 0.0551\\ 
    MCCLK & 0.0668 & 0.0428 & 0.1368 & 0.0717 & -- & --\\
    CKER & 0.0691 & 0.0433 & \underline{0.1591} & \underline{0.0851} & 0.0969 & 0.0485\\
    KACL & 0.0658 & 0.0430 & 0.1520 & 0.0812 & 0.1077 & 0.0553\\
    \midrule
    % \name w/o CL & 0.0762 & 0.0500 & 0.1439 & 0.0754 & 0.1185 & 0.0633\\
    \name & \textbf{0.0864} & \textbf{0.0567} & \textbf{0.1739} & \textbf{0.0948} & \textbf{0.1258} & \textbf{0.0662}\\
    {\em improv.\%} & {6.5\%} & {7.0\%} & 9.3\% & 11.4\% & {13.2\%} & 11.5\% \\
  \bottomrule
\end{tabular}
\end{table}

\subsection{Experimental Results}
{\bf (A) Effectiveness Comparison}.
We first compare the performance of all the methods, and the results are shown in Table~\ref{tab:performance}. From the table, we have the following observations.
%\begin{itemize}
%    \item \textbf{Our \name achieves the best results}. 
First of all, \name consistently and significantly outperforms all baselines in all cases, demonstrating its effectiveness. Specifically, compared with the best competitor, \name achieves 6.5\% - 13.2\% relative improvements. When comparing with existing KCL methods (the latter four baselines), the improvements become 9.3\% - 20.2\%.
Second, incorporating knowledge graph is helpful to improve the performance. For example, most of the knowledge-enhanced methods outperform the basic methods. %, which confirms the importance of side information in KG. 
Third, contrastive learning also helps. That is, the latter four KCL methods generally further outperform the knowledge-enhanced methods. Additionally, CL-based methods alone also perform relatively well. 
For example, SimGCL that does not use knowledge graph is the best competitor in 4 out of 6 cases (even better than KCL methods). This result indicates that an appropriate combination of knowledge graph and contrastive learning (e.g., as posed in \name) is necessary to achieve better results. %importance of generating self-supervised signals from unlabeled interactions. 

% over the strongest baselines w.r.t Recall@20 by xxx in Yelp2018, Amazon-Books and MIND, respectively. 
    
    % Overall, we attribute the performance improvements to two aspects: i) By utilizing layer-wise fusion operation of aggregated information from KG and IG, \name can learn more accurate semantic user/item representations; ii) \name construct CL views from the whole(using fused embeddings) and part(using only KG embeddings) perspectives, which captures more comprehensive self-supervised signals for \name.  %have moved it into next subsection.
    
%   \item \textbf{KG information benifit recommendation}. Most of knowledge-enhanced methods outperform BPR and NCF, which confirms the importance of side information in KG. 
   % By effectively combining the information of KG with the message aggregation process of IG, our \name can also achieve significant performance without using comparative learning(i.e., \name w/o CL), which is better than most of knowledge-enhanced methods.
   
%   \item \textbf{Contrastive learning contributes more}. The relatively superior performance achieved by SGL, SimGCL and KGCL indicates the importance of generating self-supervised signals from unlabeled interactions. 
%\end{itemize}

\begin{figure}[t]
  \centering
  \includegraphics[width=0.9\linewidth]{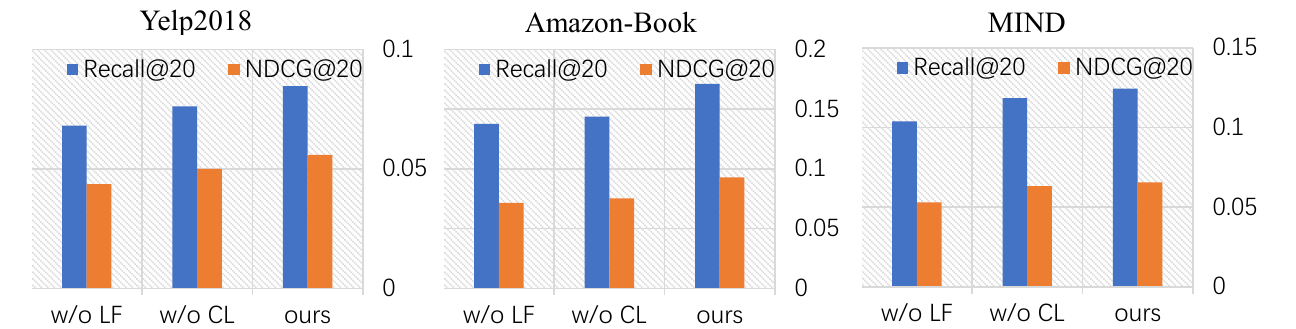}
  \caption{The performance gain of our \name. Both the layer-wise one-direction fusion and contrastive learning are important integral components.}  
  \label{fig:ablation}
  % \Description{Figure 1: fig1.}
\end{figure}

{\bf (B) Performance Gain Analysis}. 
%\subsubsection{Performance gain of \name}
In this part, we analyze the performance gain of \name. Overall, we attribute the improvements of \name to two design choices: i) our proposed one-direction layer-wise fusion operation, and ii) our proposed CL framework that constructs appropriate CL views. To this end, we conduct an ablation study, and 
%We separately verify whether the proposed model architecture(layer-wise fusion) and CL framework effective. We 
display the results in Figure~\ref{fig:ablation}, where ``w/o CL'' represents \name without CL loss, ``w/o LF'' represents \name without layer-wise fusion operation. We can observe that each part contributes to the performance gain. Note that these two design choices together form the so-called ``sweet spot'' for KCL methods.

%\subsubsection{Further Study of \name}
{\bf (C) Other Design Choices.} In order to explore the factors that affect \name, we have made a more detailed study on its implementation with other design choices. Specifically, we consider whether to apply the following operations:
% lieaner -> linear
1) project head, which introduces a learnable linear/nonlinear transformation between the representation and the CL loss~\cite{chen2020simple,chen2020big,zou2022multi}; 2) structure augmentation, which increases the randomness of CL views by randomly dropping a certain proportion of interaction edges of IG and triple edges of KG; 
%in the process of message propagation; It is noting that we do not use structural augmentation to obtain additional views(e.g., ~\cite{wu2021self,yang2022knowledge}). 
3) embedding normalization, which normalizes KG embeddings in each propagation layer; %can affects the convergence and effectiveness of model training. We present 
The results are shown in Table~\ref{tab:ablation}, where 
``w project'', ``w/o augmentation'', and ``w/o normalization'' denote implementation with project head, without structure augmentation, and without embedding normalization, respectively.
% ``default'' means the opposite of the above three implementations. 
From the table, we can observe that: 1) different from ~\cite{chen2020simple,chen2020big,zou2022multi}, introducing extra project head does not work in our paradigm, and the reason may be that our CL representations are easy to pull close with the help of layer-wise fusion; 2) structure augmentation has achieved further performance improvements in all cases;
%, which implicit that a certain degree of random edge dropping may benefits representation learning; 
3) embedding normalization in KG propagation is critical in our paradigm. 
%we argue that the embedding element value can be too large after message aggregation in Eq~\ref{eq:kg_propagate} without normalization operation, which is unfavorable to the model convergence and embedding learning.

\begin{table}[t]\centering
  \caption{The effects of different design choices of \name.}
  \label{tab:ablation}
  \begin{tabular}{ccccccc}
  \toprule
    \multirow{2}{*}{Method}
    & \multicolumn{2}{c}{Yelp2018} & \multicolumn{2}{c}{Amazon-Book} & \multicolumn{2}{c}{MIND}\\ 
    & Recall & NDCG & Recall & NDCG & Recall & NDCG \\ 
    \midrule
    w project & 0.0714 & 0.0468 & 0.1404 & 0.0735 & 0.1153 & 0.0596\\
    w/o augmentation & 0.0833 & 0.0550 & 0.1698 & 0.0927 & 0.1208 & 0.0635\\
    w/o normalization & 0.0767 & 0.0501 & 0.0590 & 0.0343 & 0.1094 & 0.0569\\ 
    \midrule
    \name & \textbf{0.0864} & \textbf{0.0567} & \textbf{0.1739} & \textbf{0.0948} & \textbf{0.1258} & \textbf{0.0662}\\
  \bottomrule
\end{tabular}
\end{table}

\begin{table}[t]\centering
  \caption{Performance of using NGCF~\cite{wang2019neural} as the IG backbone and using KGAT~\cite{wang2019kgat} as the KG backbone. Our framework is still better.}
  \label{tab:other_backbones}
  \begin{tabular}{ccccccc}
  \toprule
    \multirow{2}{*}{Method}
    & \multicolumn{2}{c}{Yelp2018} & \multicolumn{2}{c}{Amazon-Book} & \multicolumn{2}{c}{MIND}\\ 
    & Recall & NDCG & Recall & NDCG & Recall & NDCG \\ 
    \midrule
    NGCF & 0.0587 & 0.0376 & 0.1139 & 0.0574 & 0.0961 & 0.0483\\
    KGAT & 0.0675 & 0.0432 & 0.1390 & 0.0739 & 0.0907 & 0.0442\\
    KGCL & 0.0756 & 0.0493 & 0.1496 & 0.0793 & 0.1073 & 0.0551\\
    SimGCL & 0.0811	& 0.0530 & 0.1579 &0.0848 & 0.1111 & 0.0594 \\ 
    \midrule
    \name w NGCF & 0.0794 & 0.0470 & \textbf{0.1645} & \textbf{0.0913} & 0.1082 & 0.0547\\
    \name w KGAT & \textbf{0.0827} & \textbf{0.0545} & 0.1619 & 0.0895 & \textbf{0.1240} & \textbf{0.0656} \\
  \bottomrule
\end{tabular}
\end{table}

{\bf (D) Different GNN Backbones.} %\name with other backbones
%When separately encoding the graphs, our method follows a . 
In order to further verify the proposed paradigm, we further build our method upon other GNN backbones to demonstrate that \name can be a GNN-agnostic CL paradigm for knowledge-enhanced recommendation. Specifically, we replace our KG and IG encodings with other choices and keep the other part unchanged. The results are reported in Table~\ref{tab:other_backbones}, where ``\name w NGCF'' and ``\name w KGAT'' represent that we utilize NGCF to encode IG and utilize KGAT to encode KG, respectively. As we can see, both variants have achieved great improvements compared with the backbone model alone, and is even better than the two best competitors (KGCL and SimGCL).
% as indicated by Table~\ref{tab:performance}.

% In \name, we build our method upon the LightGCN model. However,it is unclear if our method is also applicable to other models.
% To answer this question, we next investigate \name on the NGCF model. 
% We utilize NGCF as the GNN encoder, and then verify the effectiveness of each module and their combination on three datasets: Gowalla-20\%, Ylep-20\%, and Amazon-20\%. The results are reported in Table~\ref{tab:NGCF}, where ``NGCF w SD'', ``NGCF w EP'', and ``NGCF w Both'' represent the NGCF with our proposed structure denoising module, embedding space perturbation module, and both, respectively. 
% As we can see from the table, both of the proposed structure denoising module and the embedding space perturbation module improve the performance of NGCF, although the improvement is smaller than that of LightGCN. The possible reason is that  LightGCN is easier to optimize with a simpler structure.

\begin{figure}
  \centering
  \includegraphics[width=0.8\linewidth]{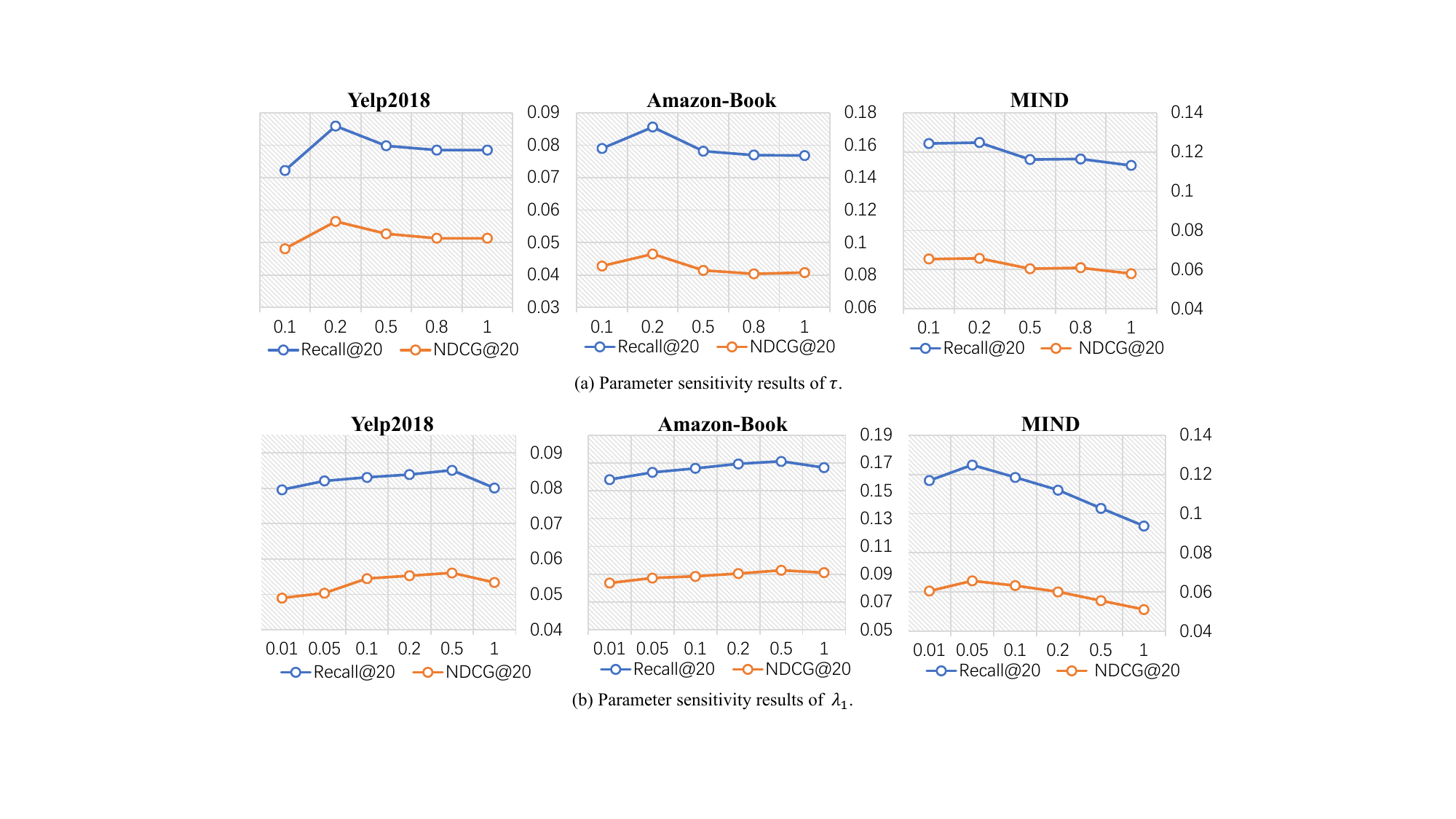}
  \caption{Parameter sensitivity results of $\tau$ and $\lambda_{1}$.}
  \label{fig:parameter}
\end{figure}

% \begin{figure}
%   \centering
%   \includegraphics[width=0.9\linewidth]{fig/tau.pdf}
%   \caption{Parameter sensitivity results of $\tau$.}
%   \label{fig:tau}
%   % \Description{tau}\label{fig:tau}
% \end{figure}

% \begin{figure}
%   \centering
%   \includegraphics[width=0.9\linewidth]{fig/lambda.pdf}
%   \caption{Parameter sensitivity results of $\lambda_{1}$.}
%   \label{fig:lambda}
%   % \Description{lambda_1}\label{fig:lambda}
% \end{figure}

\begin{table}[t]\centering
  \caption{The training time comparison.}
  \label{tab: running time}
  %\footnotesize
  \begin{tabular}{ccrrr}
  \toprule
    Methods & Component & Yelp2018(s) & Amazon-Book(s) & MIND(s) \\
  \midrule
	\multirow{3}{*}{LightGCN} 
	& Single epoch time & 12.94 & 8.33 & 82.59 \\
	& Total training epochs & 180 & 207 & 52 \\
	& Total training time & 2,329.20 ($1\times$) & 1,724.31 ($1\times$) & 4,294.68 ($1\times$)\\
  	\midrule
	\multirow{3}{*}{KGCL} 
	& Single epoch time & 383.82 & 196.11 &  1103.31\\
	& Total training epochs & 30 & 22 & 18\\
	& Total training time & 11,514.60 ($4.94\times$) & 4,314.42 ($2.50\times$) & 19,859.58 ($4.62\times$)\\
  	\midrule
  	\multirow{3}{*}{\name} 
	& Single epoch time & 144.33 & 232.06 & 324.55\\
	& Total training epochs & 22 & 6 & 2\\
	& Total training time & 3175.26 ($1.36\times$) & 1392.36 ($0.81\times$) & 649.10 ($0.15\times$)\\
  \bottomrule
\end{tabular}
\end{table}

{\bf (E) Parameter Sensitivity}.
Next, we analyze the impact of two hyper-parameters (i.e., $\tau$, and $\lambda_{1}$) in \name. 
$\tau$ in Eq.~\ref{eq:cl_user} is the softmax temperature coefficient. We fix other hyper-parameters and then tune it in the interval [0.1, 1.0]. As shown in Figure~\ref{fig:parameter}(a), the model performance is relatively stable when $\tau$ varies in a large range. % and the best performance is achieved when $\tau$ is set as 0.2. 
$\lambda_1$ in Eq.~\ref{eq:all_loss} is a regularization coefficient controlling the relative weight of contrastive learning loss. We fine-tuned $\lambda_{1}$ in the interval [0.01, 1.0] as discussed in the experimental setup. The results are shown in  Figure~\ref{fig:parameter}(b). As we can see, the performance of the model improves with the increase of the $\lambda_{1}$ value at first, and then starts to decrease after a certain value. Generally speaking, the performance of \name stays stable in a relatively wide range w.r.t. the two hyper-parameters. % The best values on the Yelp2018, Amazon-Book, and MIND are 0.5, 0.5, and 0.05, respectively.

{\bf (F) Efficiency}.
Finally, we evaluate the efficiency aspect of the proposed method.
We compare \name with LihgtGCN and KGCL. We choose these two baselines because LightGCN serves as a GNN-based IG graph encoder famous for its lightweight design and efficiency, and KGCL serves as the state-of-the-art method for KCL methods and it performs relatively stable. The results are shown in Table~\ref{tab: running time} which are collected on an Intel(R) Xeon(R) Silver 4110 CPU and a GeForce RTX 3080Ti GPU.
%As shown in Table~\ref{tab: running time}, we count the training time of every single epoch, the number of training epochs to converge, and the total training time for all methods. 
We can observe that \name's total training time is much shorter than KGCL, and even shorter than the lightweight method LightGCN on Amazon-Book and MIND, which is mainly due to its faster convergence speed.

%the training time per epoch of our method is surprisingly reduced compared to KGCL. The reason is that there is no reliance on graph structure augmentation to construct comparative learning views in our method. 
%In addition, \name's total training time is much shorter than KGCL, and even shorter than that of baseline LightGCN on Amazon-Book and MIND, which mainly due to its faster convergence speed. In conclusion, extensive experiments have proved that our method is efficient and effective.

\section{Conclusions}
In this paper, we have proposed a new contrastive learning framework \name for knowledge-enhanced recommendation. Different from the existing KCL methods, the key design of \name is to allow one-direction fusion from knowledge graph encoding to interaction graph encoding, and then use both as contrastive views.
Through such simple but effective design,
we conduct extensive experiments and demonstrate that \name achieves 6.5\% - 13.2\% relative improvements compared the best competitor among 16 baselines, and meanwhile requires much shorter training time compared with the state-of-the-art KCL method KGCL.

\bibliographystyle{plainnat}
\bibliography{recommend}

\end{document}